\documentclass[aps,pra,twocolumn,floatfix,showpacs]{revtex4}
\usepackage[dvips]{color}
\usepackage{amsmath,amssymb,amsfonts,bm,graphicx,latexsym}
\usepackage[T1]{fontenc}
\usepackage[hypertex,breaklinks=true,colorlinks=false]{hyperref}

\newcommand{\ket}[1]{|#1 \rangle}

\newcommand{\Kcal}{{\cal K}}

\newcommand{\rv}{{\bm{r}}}

\newcommand{\pv}{\bm{p}}
\newcommand{\qv}{\bm{q}}
\newcommand{\Kv}{{\bm{K}}}
\newcommand{\piv}{\bm{\pi}}

\newcommand{\Av}{\bm{A}}
\newcommand{\Bv}{\bm{B}}

\newcommand{\ah}{\hat{a}}

\newcommand{\Hh}{\hat{H}}

\newcommand{\Nh}{\hat{N}}

\newcommand{\Psih}{\hat{\Psi}}
\newcommand{\rhoh}{\hat{\rho}}
\newcommand{\nh}{\hat{n}}

\newcommand{\Vt}{\tilde{V}}
\newcommand{\Zbb}{\mathbb{Z}}

\newcommand{\tot}{\mathrm{tot}}

\newcommand{\PRL}[3]{Phys. Rev. Lett. {\bf #1}, \href{http://link.aps.org/abstract/PRL/v#1/e#2}{#2} (#3)}
\newcommand{\PRLp}[3]{Phys. Rev. Lett. {\bf #1}, \href{http://link.aps.org/abstract/PRL/v#1/p#2}{#2} (#3)}
\newcommand{\PRA}[3]{Phys. Rev. A {\bf #1}, \href{http://link.aps.org/abstract/PRA/v#1/e#2}{#2} (#3)}

\newcommand{\PRAR}[3]{Phys. Rev. A {\bf #1}, \href{http://link.aps.org/abstract/PRA/v#1/e#2}{#2} (R) (#3)}
\newcommand{\PRB}[3]{Phys. Rev. B {\bf #1}, \href{http://link.aps.org/abstract/PRB/v#1/e#2}{#2} (#3)}
\newcommand{\PRBp}[3]{Phys. Rev. B {\bf #1}, \href{http://link.aps.org/abstract/PRB/v#1/p#2}{#2} (#3)}
\newcommand{\PRBR}[3]{Phys. Rev. B {\bf #1}, \href{http://link.aps.org/abstract/PRB/v#1/e#2}{#2} (R) (#3)}

\newcommand{\RMP}[3]{Rev. Mod. Phys. {\bf #1}, \href{http://link.aps.org/abstract/RMP/v#1/e#2}{#2} (#3)}
\newcommand{\arXiv}[1]{arXiv:\href{http://arxiv.org/abs/#1}{#1}}

\begin{document}

\title{
Devil's staircases in synthetic dimensions and gauge fields
}
\author{Takeshi Y. Saito}
\affiliation{Department of Physics, University of Tokyo, 7-3-1 Hongo, Bunkyo-ku, Tokyo 113-0033, Japan}
\author{Shunsuke Furukawa}
\email{furukawa@cat.phys.s.u-tokyo.ac.jp}
\affiliation{Department of Physics, University of Tokyo, 7-3-1 Hongo, Bunkyo-ku, Tokyo 113-0033, Japan}
\date{\today}


\begin{abstract}
We study interacting bosonic or fermionic atoms in a high synthetic magnetic field in two dimensions 
spanned by continuous real space and a synthetic dimension.
Here, the synthetic dimension is provided by hyperfine spin states, 
and the synthetic field is created by laser-induced transitions between them. 
While the interaction is short-range in real space, 
it is long-range in the synthetic dimension in sharp contrast with fractional quantum Hall systems. 
Introducing an analog of the lowest-Landau-level approximation valid for large transition amplitudes, 
we derive an effective one-dimensional lattice model, 
in which density-density interactions turn out to play a dominant role. 
We show that in the limit of a large number of internal states, the system exhibits a cascade of crystal ground states, which is known as devil's staircase, 
in a way analogous to the thin-torus limit of quantum Hall systems. 
\end{abstract}

\maketitle


\section{Introduction} \label{sec:intro}

Laser-induced gauge fields in ultracold atomic systems have attracted considerable attention in recent years \cite{Goldman13,Dalibard11}. 
By optically coupling internal states of atoms, a uniform magnetic field \cite{Lin09} and a spin-orbit coupling \cite{Lin11,Wang12,Zhai12}
have been engineered in quantum gases. 
Furthermore, a Hofstadter Hamiltonian \cite{Harper55,Hofstadter76}, 
in which a uniform flux pierces through each plaquette of a square lattice, has been experimentally realized 
by using laser-assisted tunneling in optical lattices \cite{Aidelsburger13,Miyake13,Aidelsburger15,Kennedy15}. 
By using these techniques, we can expect to emulate quantum Hall states and other topological states of matter in highly controlled atomic systems
and elucidate yet unexplored effects of quantum statistics and strong correlations on those states \cite{Goldman16,Bloch12}. 
For interacting scalar Bose gases in high synthetic magnetic fields, which have no analog in solid-state physics, 
a variety of quantum Hall states have been predicted to appear \cite{Cooper08_review}, 
including a bosonic Laughlin state \cite{Laughlin83,Wilkin98} and non-Abelian Read-Rezayi states \cite{Read99,Cooper01}. 

\begin{figure}
\begin{center}\includegraphics[width=0.47\textwidth]{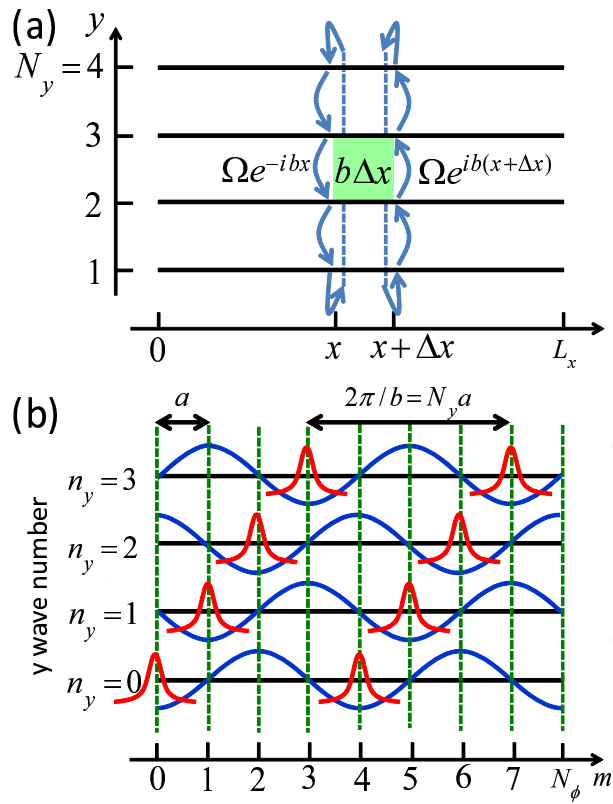}\end{center}
\caption{(color online) (a) A system of atoms confined in continuous 1D space and having $N_y$ hyperfine spin states labeled by $y=1,2,\dots, N_y$ [see Eq.~\eqref{eq:Hkin}]. 
Transitions with a spatially-dependent phase factor $e^{ibx}$ are induced between these internal states. 
For an interval of length $\Delta x$, a magnetic flux of $b\Delta x$ pierces between the neighboring internal states. 
(b) By performing the Fourier transformation in the $y$ direction, the kinetic part of the Hamiltonian decouples into components 
with different values of the wave number $k_y=2\pi n_y/N_y$ in the $y$ direction [see Eqs.~\eqref{eq:Hkin_ky} and \eqref{eq:h_ky}]. 
For each $k_y$, an effective cosine potential emerges with shifted locations of minima, allowing a description in terms of Wannier orbitals. 
These cosine potentials may also be viewed as spatially-dependent optically dressed levels.
The Wannier orbitals are labeled by a serial number $m$ [see Eq.~\eqref{eq:m_serial}] from left to right, forming a 1D lattice.}
\label{fig:system}
\end{figure}

Controllability of gauge fields can further be facilitated by exploiting internal states of atoms as a {\it synthetic} dimension \cite{Boada12, Celi14}. 
In a two-dimensional (2D) lattice spanned by a 1D optical lattice and a synthetic dimension, 
a uniform gauge flux can be created by suitable laser-induced transitions between internal states, 
allowing the realization of a Hofstadter model \cite{Celi14,Mancini15,Stuhl15}. 
The length in the synthetic dimension is tunable and can be as large as $6$ for $^{173}$Yb atoms \cite{Taie12,Pagano14} and $10$ for $^{40}$K atoms \cite{Krauser12}. 
This scheme is flexible in designing different boundary conditions, 
offering advantages in studying topological states of matter. 
Naturally sharp edges in the synthetic dimension leads to the formation of chiral edge states, as observed experimentally \cite{Mancini15,Stuhl15}. 
A periodic boundary condition in the synthetic dimension can be realized by further coupling the two edges by lasers, 
allowing the formation of a Hofstadter-like fractal spectrum \cite{Celi14}. 
An attractive proposal of realizing a four-dimensional quantum Hall effect by using a synthetic fourth dimension has also been made \cite{Price15}. 

The theoretical proposals of Refs.~\cite{Celi14,Price15} have mainly focused on non-interacting systems. 
Given the emergence of quasi-flat bands in the Hofstadter model, the interaction is expected to play a significant role 
unless the Fermi energy lies between the bands. 
Notably, while the interaction is short-range in real space, 
it is long-range in the synthetic dimension---a spin-independent interaction gives a leading part of the interaction in many atomic systems. 
This anisotropic nature of the interaction sharply contrasts with fractional quantum Hall (FQH) systems, 
and has been shown to lead to density waves \cite{Barbarino15,Zeng15_pumping,Hurst16}, fractional charge pumping \cite{Zeng15_pumping,Taddia17}, 
helical liquids \cite{Barbarino15, Petrescu15, Cornfeld15, Hurst16, Barbarino16,Strinati17}, supersolids, pair superfliuds \cite{Bilitewski16}, 
and other interesting phenomena \cite{Natu15,Ghosh15,Uchino16}. 

In this paper, we study a system of interacting bosonic or fermionic atoms in a high synthetic magnetic field 
in two dimensions spanned by continuous real space and a synthetic dimension as shown in Fig.~\ref{fig:system}(a). 
The synthetic field is created by laser-induced transitions between internal states. 
In contrast to the settings considered previously \cite{Celi14, Barbarino15, Zeng15_pumping}, 
we do not introduce an optical lattice in 1D real space but allow atoms to move continuously in this direction. 
This simplicity of the system allows us to introduce an analog of the lowest-Landau-level (LLL) basis 
as used in the studies of the FQH effect \cite{Yoshioka84,Haldane85}. 
We focus on the limit when the transition amplitudes between internal states are sufficiently large, 
and derive an effective 1D lattice model in the LLL-like basis, 
in which particle hopping is significantly suppressed and density-density interactions play a dominant role. 
This effective Hamiltonian is analogous to the thin-torus limit of quantum Hall systems \cite{Tao83,Seidel05,Bergholtz07,Rotondo16}. 
In the limit of a large number of internal states, we show the emergence of 
a complete devil's staircase---a crystal ground state (GS) (i.e., a density wave state) appears for every rational filling with increasing the chemical potential. 
We also analyze the stability of the crystal GSs in the case when the number of internal states is small, 
and discuss connections of our results with Refs.\ \cite{Barbarino15, Zeng15_pumping}. 

We note that the properties of the present system is quite sensitive to the boundary condition in the synthetic dimension. 
For open boundary conditions in the synthetic dimension, if the transition amplitudes between internal states are sufficiently large, 
one can first solve the atomic problem and restrict ourselves to the lowest-energy dressed state; 
since the energy of this state does not depend on the position in real space \cite{Comment_dressed}, 
the system can be treated as a 1D single-component gas \cite{Celi14, Bilitewski16}. 
In contrast, we here impose a periodic boundary condition in the synthetic dimension as in Refs.\ \cite{Barbarino15, Zeng15_pumping}. 
In this case, the optically dressed levels are spatially-dependent and sometimes cross with each other 
[as seen in the effective cosine potentials in Fig.~\ref{fig:system}(b)]. 
This structure of dressed levels makes the problem more complex and gives rise to the formation of crystal GSs. 

The rest of the paper is organized as follows. 
In Sec.~\ref{sec:model_Heff}, we introduce our model, 
and derive an effective lattice Hamiltonian within the LLL-like approximation 
valid for large transition amplitudes between internal states. 
In Sec.~\ref{sec:gs}, we analyze the GS phase diagram of the model by using the effective Hamiltonian, 
and show the appearance of devil's staircases in the limit of a large number of internal states. 
We also analyze the stability of crystal GSs when the number of internal states is small. 
We conclude the paper in Sec.~\ref{sec:summary}. 
In Appendix, we discuss a related problem in the continuum limit. 

\section{Model and effective Hamiltonian} \label{sec:model_Heff}

\newcommand{\kin}{\mathrm{kin}}
\newcommand{\interact}{\mathrm{int}}
\newcommand{\drm}{\mathrm{d}}


We consider a system of interacting bosonic or fermionic atoms confined in 1D space in the $x$ direction 
and having $N_y$ hyperfine spin states labeled by $y=1,2,\dots,N_y$. 
Transitions with an $x$-dependent phase factor $e^{ibx}$ are induced 
between the internal states as in Fig.~\ref{fig:system}(a). 
This setup realizes a high synthetic magnetic field as proposed and experimentally demonstrated in Refs.~\cite{Celi14,Mancini15,Stuhl15}. 
If the internal states labeled by $y$ correspond to the spin-$F$ sublevels $m=-F,-F+1,\dots,+F$, 
Raman lasers can realize such transitions with an amplitude 
$\Omega_m=(\Omega_R/2)\sqrt{F(F+1)-m(m+1)}$ between the sublevels $m$ and $m+1$ \cite{Celi14}. 
A transition between $m=\pm F$ can further be induced 
by using additional Raman and radio-frequency transitions \cite{Celi14}. 
In our analysis, we focus on the case of $y$-independent transition amplitudes $\Omega(>0)$ 
as the emergence of crystal GSs can be elucidated most clearly in this case. 
We note that for $F=1$, one naturally has $\Omega_{-1}=\Omega_0$ 
and the transition amplitude between $m=\pm 1$ can be tuned separately, 
enabling the realization of a uniform amplitude $\Omega$. 

The kinetic part of the Hamiltonian is given by
\begin{equation}\label{eq:Hkin}
\begin{split}
 \Hh_\kin = \int _0^{L_x}  & d x   \sum _{y=1}^{N_y} 
 \bigg\{ \Psih^\dag (x,y) \left( -\frac{\hbar^2\partial_x^2}{2M} \right) \Psih (x,y) \\
 &-\Omega \left[e^{ibx} \Psih^{\dag}(x,y+1)\Psih(x,y)+ \mathrm{h.c.}\right] \bigg\} ,
\end{split}
\end{equation}
where $L_x$ is the length in the $x$ direction and $M$ is the particle's mass \cite{Budich17}. 
Here, $\Psi(x,y)$ is a bosonic or fermionic field operator for the $y$-th component satisfying the commutation relations
\begin{equation}
\begin{split}
 &[\Psih(x,y),\Psih^\dagger (x',y')]_{B/F} = \delta(x-x')\delta_{y,y'},\\
 &[\Psih(x,y),\Psih (x',y')]_{B/F} =0, 
\end{split}
\end{equation}
where $[\hat{X},\hat{Y}]_B\equiv \hat{X}\hat{Y}- \hat{Y}\hat{X}$ and $[\hat{X},\hat{Y}]_F\equiv \hat{X}\hat{Y}+ \hat{Y}\hat{X}$ for the bosonic and fermionic cases, respectively. 
We impose periodic boundary conditions
\begin{equation}
 \Psih(L_x,y)=\Psih(0,y),~\Psih(x,N_y+1)=\Psih(x,1). 
\end{equation}
We also assume that the phase factor $e^{ibx}$ of the laser-induced transitions is compatible with this periodicity, i.e., 
$bL_x=2\pi N_x$, where $N_x$ is an integer. 
The total number of flux quanta piercing through the system is $N_\phi=N_x N_y$. 
We note that the periodic boundary condition in the $x$ direction is only for the convenience of 
our analysis---when $L_x$ is sufficiently large, 
the bulk properties does not depend crucially on the boundary condition in this direction. 
In contrast, $N_y$ is limited by the number of atomic internal states that can be coupled by lasers, 
and thus the boundary condition in the $y$ direction is crucial. 
While we keep $N_y$ finite in the derivation of the effective Hamiltonian, 
a remarkable simplification is achieved by subsequently taking the limit $N_y\to\infty$ in the derived effective Hamiltonian, as we discuss later. 

The inter-particle interaction in this system has a highly anisotropic nature---it is long-range and non-decaying in the synthetic dimension. 
For simplicity, we consider a spin-independent contact interaction between atoms, 
which gives a leading part of the interaction in many ultracold atomic systems.  
The interaction Hamiltonian is then given by
\begin{equation}\label{eq:Hint}
 \Hh_\interact = \frac{g}{2} \int_0^{L_x} \! d x~ :\! [\rhoh_\tot(x)]^2 \!:, 
\end{equation}
where $g$ is the interaction strength, colons indicate the normal ordering, 
and $\rho_\tot (x)$ is the total density operator defined by
\begin{equation}\label{eq:rho_tot}
 \rhoh_\tot (x) = \sum_{y=1}^{N_y} \Psih^\dagger (x,y) \Psih (x,y).
\end{equation}
The total Hamiltonian of the system is given by the sum of the kinetic and interaction Hamiltonians: 
$ \Hh = \Hh_\kin + \Hh_\interact $. 
When $\Omega=0$, this Hamiltonian reduces to the Hamiltonian of an interacting Bose or Fermi gas with SU$(N_y)$ symmetry. 

We now derive an effective Hamiltonian which is useful for discussing the GS for a large transition amplitude $\Omega$. 
To this end, we first perform the Fourier expansion of the field operator in the $y$ direction:
\begin{equation}
 \Psih(x,y)=\frac{1}{\sqrt{N_y}} \sum_{k_y} e^{ik_y y} \Psih (x,k_y)
\end{equation}
with
 $k_y = 2\pi n_y/N_y ~(n_y=0,1,2,\dots ,N_y-1)$.
The kinetic Hamiltonian is then decoupled into components with different $k_y$ as
\begin{equation}\label{eq:Hkin_ky}
 \Hh_\kin =\sum_{k_y} \int_0^{L_x} \! d x~ \Psih^\dagger (x, k_y) \Kcal_{k_y} \Psih(x,k_y),
\end{equation}
where
\begin{equation}\label{eq:h_ky}
 \Kcal_{k_y} = - \frac{\hbar^2\partial_x^2}{2M} -2\Omega \cos(bx-k_y) .
\end{equation}
Interestingly, the single-particle Hamiltonian $\Kcal_{k_y}$ for the wave number $k_y$ 
has the same form as the Hamiltonian of a particle living in a cosine potential 
whose locations of minima are displaced by $k_y/b=2\pi n_y/(N_y b)\equiv n_y a$ as shown in Fig.~\ref{fig:system}(b). 
In analogy with an optical lattice, we introduce the ``recoil energy'' $E_\mathrm{r}\equiv \hbar^2 (b/2)^2/(2M)$, 
which gives the scale of the kinetic energy of a particle in the effective cosine potential.  

In the limit of a large transition amplitude $4\Omega \gg E_\mathrm{r}$, 
the cosine potential in Eq.~\eqref{eq:h_ky} becomes deep, 
and it is useful to focus on Wannier orbitals which are localized around the minima of this potential. 
For given $k_y=2\pi n_y /N_y$, such minima are located at
\begin{equation}
 x=\frac{2\pi n_x+k_y}{b} = (N_yn_x+n_y)a, ~n_x=0,1,\dots,N_x-1. 
\end{equation}
Considering these minima for all possible $n_y$, we find that they appear with the spacing of $a$ as in Fig.~\ref{fig:system}(b). 
It is thus useful to introduce the serial number
\begin{equation}\label{eq:m_serial}
m=N_yn_x+n_y=0,1,\dots,N_\phi-1
\end{equation}
which labels them from left to right; this can be used as a ``site'' label for an effective 1D lattice model. 
The field operator is now expanded as
\begin{equation}\label{eq:Psi_w}
 \Psih(x,y)=\frac{1}{\sqrt{N_y}} \sum_{m=0}^{N_\phi-1} e^{i\frac{2\pi m}{N_y}y} \phi(x-ma) \ah_m, 
\end{equation}
where $\phi(x)$ is the wave function of the Wannier orbital, 
and $\ah_m$ is the annihilation operator at the $m$-th ``site'' satisfying the commutation relations 
\begin{equation}\label{eq:ah_comm}
 [\ah_m,\ah_{m'}^\dagger]_{B/F}=\delta_{mm'},~~[\ah_m,\ah_{m'}]_{B/F}=0.
\end{equation}
Equation \eqref{eq:Psi_w} is an expansion over the $N_\phi$ single-particle states 
and can be viewed as an analog of the LLL approximation for FQH systems 
(see Appendix for the description of a related problem in the continuum limit). 
By substituting Eq.~\eqref{eq:Psi_w} into Eq~\eqref{eq:Hkin} and \eqref{eq:Hint}, 
we obtain the effective 1D lattice model with 
\begin{align}
 &\Hh_\kin = -\sum_{m,m'=0}^{N_\phi-1}  J_{mm'} \ah_m^\dagger \ah_{m'}, \label{eq:Kin_Jaa}\\
 &\Hh_\interact = \frac12 \sum_{ \{m_j\} }  V_{m_1m_2m_3m_4} \ah_{m_1}^\dagger \ah_{m_2}^\dagger \ah_{m_3} \ah_{m_4} \label{eq:Hint_Vaaaa},
\end{align}
where
\begin{align}
 &J_{mm'} = -\delta_{mm'}^{(N_y)} \int_0^{L_x} \!\! d x ~\phi(x-ma) \Kcal_{k_y} \phi(x-m'a), \label{eq:Jmm} \\
 &V_{m_1m_2m_3m_4} = g \delta_{m_1m_4}^{(N_y)} \delta_{m_2m_3}^{(N_y)} \int_0^{L_x} \!\! d x \prod_{j=1}^4 \phi(x-m_j a) , \label{eq:Vmmmm}
\end{align}
and $\delta_{mm'}^{(N_y)}$ is the Kronecker delta of period $N_y$, i.e., 
\begin{equation}\label{eq:Kronecker_p}
 \delta_{mm'}^{(N_y)}=
 \begin{cases} 1~~(m\equiv m'~\text{mod}~N_y); \\ 0~~(\text{otherwise}). \end{cases}
\end{equation} 

\begin{figure}[b]
\begin{center}\includegraphics[width=0.47\textwidth]{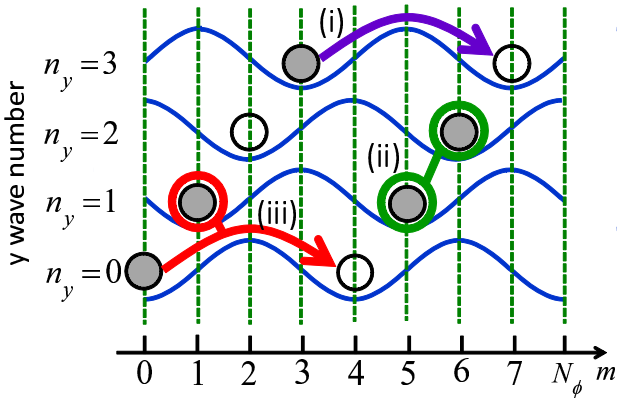}\end{center}
\caption{(color online) Processes in the effective Hamiltonian [see Eqs.\ \eqref{eq:Hint_Vaaaa} and \eqref{eq:Hkin_Jaa}]: 
(i) hopping $\ah_{m+N_y}^\dagger \ah_m$, (ii) a density-density interaction $\nh_m \nh_{m'}$,  and
(iii) correlated hopping $\nh_{m'} \ah_{m+N_y}^\dagger \ah_{m}$.
Shaded and empty circles indicate the presence and absence of a particle, respectively.}
\label{fig:Heff}
\end{figure}

The kinetic Hamiltonian \eqref{eq:Kin_Jaa} now gives hopping terms among ``sites''. 
As seen in Eq.~\eqref{eq:Jmm}, hopping occurs only when the distance $m-m'$ between ``sites'' is an integer multiple of $N_y$  
because of the conservation of the momentum in the $y$ direction [see also (i) in Fig.~\ref{fig:Heff}].  
Similar to the case of an optical lattice, we can restrict the hopping to the shortest distance $m-m'=\pm N_y$ 
as the hopping amplitudes for further distances are much smaller
(we note that terms with $m=m'$ are also present but can be absorbed into the definition of the chemical potential).
We then obtain 
\begin{equation}\label{eq:Hkin_Jaa}
 \Hh_\kin = -J \sum_m (\ah_{m+N_y}^\dagger \ah_m + \mathrm{h.c.} ),
\end{equation}
where $J$ is given through an exact solution of the bandwidth by \cite{Zwerger03}
\begin{equation}\label{eq:J_Ny}
\begin{split}
J
&=\frac{4}{\sqrt{\pi}} E_\mathrm{r} \left( \frac{4\Omega}{E_\mathrm{r}} \right)^{3/4} \exp \left[ -2 \left( \frac{4\Omega}{E_\mathrm{r}}\right)^{1/2} \right] \\
&=\frac{4}{\sqrt{\pi}} E_\mathrm{r} \left( \frac{N_y a}{\pi \ell} \right)^3 \exp\left[ -2 \left(\frac{N_y a}{\pi \ell} \right)^2 \right].
\end{split}
\end{equation}
Here, $\ell$ is an analog of a magnetic length and is defined as the width of the Wannier orbital [see Eq.~\eqref{eq:w_Gauss} below]. 

The interaction Hamiltonian \eqref{eq:Hint_Vaaaa} with Eq.~\eqref{eq:Vmmmm} contains density-density interactions $\nh_{m_1} \nh_{m_2}$ 
as well as correlated hopping processes $\nh_{m_1} \ah_{m_2}^\dagger \ah_{m_3}$ [see (ii) and (iii) in Fig.~\ref{fig:Heff}], 
where $\nh_m=\ah_m^\dagger \ah_m$ is the number operator at the $m$-th ``site''.  
Here, hopping occurs only for distances which are integer multiples of $N_y$ because of the $y$-momentum conservation. 
The coefficients \eqref{eq:Vmmmm} can be calculated by approximating the Wannier orbital by a Gaussian
\begin{equation}\label{eq:w_Gauss}
 \phi(x)=\frac1{\pi^{1/4}\ell^{1/2}} e^{-x^2/(2\ell^2)},  ~~\ell=\sqrt{\frac{\hbar}{(2M\Omega)^{1/2} b}}. 
\end{equation}
By substituting Eq.\ \eqref{eq:w_Gauss} into Eq.\ \eqref{eq:Vmmmm} and taking the limit $L_x\to\infty$, we obtain
\begin{equation}\label{eq:V_Gauss}
 V_{m_1m_2m_3m_4}=\frac{g\delta_{m_1m_4}^{(N_y)} \delta_{m_2m_3}^{(N_y)} }{\sqrt{2\pi}\ell} 
 \exp\left[ -\frac{a^2}{8\ell^2} \sum_{i<j} (m_i-m_j)^2 \right].
\end{equation}
Equations \eqref{eq:Hint_Vaaaa} and \eqref{eq:Hkin_Jaa} with Eqs.\ \eqref{eq:J_Ny} and \eqref{eq:V_Gauss} 
give the full expression of our effective Hamiltonian. 

We now consider the limit of a large number of internal states ($N_y\to\infty$), 
where the effective Hamiltonian can further be simplified. 
In this limit, direct hopping as well as correlated hopping processes are suppressed 
because the minimal distance of the hopping is $N_y$ 
and the hopping amplitude drops rapidly as a function of the distance. 
Thus, only the density-density interactions survive, leading to a very simple Hamiltonian 
\begin{equation}\label{eq:H_Vnn}
 \Hh=\sum_{m} \sum_{r\ge 0} \frac{1}{2^{\delta_{r,0}}} V_r :\nh_m \nh_{m+r} :, 
\end{equation}
where 
\begin{equation}\label{eq:Vr}
 V_r 
 = \frac1{2} \left( V_{0rr0}+V_{r00r} \right)
 = \frac{g}{\sqrt{2\pi}\ell} \exp \left[-\frac{a^2}{2\ell^2} r^2 \right].
\end{equation}
The Hamiltonian \eqref{eq:H_Vnn} is readily diagonal in the Fock basis $\ket{ \{n_m\} }$ and can be analyzed in a much simpler manner than the original Hamiltonian. 
Physically, the limit $N_y\to\infty$ corresponds to the case of the continuous $y$ coordinate, 
and the suppression of hopping in this limit can be understood from the formation of flat Landau levels in continuous 2D space. 
Indeed, within the LLL approximation, the same effective Hamiltonian as Eq.\ \eqref{eq:H_Vnn} can be obtained 
for a long-range interaction in the continuous $y$ coordinate, as described in Appendix. 
In the next section, we see that the repulsive interactions in this Hamiltonian stabilize a cascade of crystal GSs. 
We note that since the width of the Gaussian interaction potential \eqref{eq:Vr} in units of the lattice spacing is given by
\begin{equation}
 \frac{\ell}{a} = \frac{N_y (\hbar b)^{1/2}}{2\pi (2M\Omega)^{1/4}}, 
\end{equation}
we must tune $\Omega$ or $b$ at the same time to keep this width finite in the limit $N_y\to \infty$. 
In this case, the condition $E_\mathrm{r}\ll 4\Omega$ for the tight-binding approximation is automatically fulfilled 
since $4\Omega/E_\mathrm{r}=(N_y a/\pi \ell)^4\to \infty$. 

\section{Ground-state phase diagram} \label{sec:gs}

\subsection{Emergence of crystal  ground states for $N_y\to\infty$}

We investigate the GS phase diagram of the system by using the effective Hamiltonian. 
We first consider the limit $N_y\to\infty$, when the effective Hamiltonian \eqref{eq:H_Vnn} consists only of density-density interactions, 
and discuss the phase diagram in the plane spanned by $\ell/a$ and the chemical potential $\mu$. 
This effective Hamiltonian is analogous to the one studied by 
Hubbard \cite{Hubbard78}, and Pokrovsky and Uimin \cite{Pokrovsky78} (see also Ref.\ \cite{Bergholtz07,Rotondo16, Bak82, Burkov83, Burnell09} for related studies). 
For fermions or hard-core bosons, it was shown that if the interaction potential $V_r$ is convex, i.e., $V_{r+1}+V_{r-1}>2V_r$ for any $r\ge 2$, 
the Hamiltonian \eqref{eq:H_Vnn} exhibits a complete devil's staircase---a crystal GS emerges for every rational filling $\nu=N/N_\phi$ 
with increasing the chemical potential $\mu$ \cite{Hubbard78,Bak82}. 
For the Gaussian interaction potential \eqref{eq:Vr}, 
the condition of convexity is satisfied for $\ell/a<\sigma_0$, where $\sigma_0\simeq 1.957$ is obtained by solving $V_3+V_1=2V_2$ for $\ell/a$. 
If this condition is satisfied, the GS for every rational filling $\nu=p/q$ 
(with $p$ and $q$ being relatively coprime) is periodic with period $q$ \cite{Hubbard78}, 
is unique up to global lattice translations \cite{Burkov83}, 
and can be constructed through the procedure described in Refs.~\cite{Hubbard78, Burnell09}. 
For $\nu=1/q$, for example, particles are arranged with a mutual separation of $q$ lattice sites, which we denote by $\overline{q}$. 
The GS at $\nu=2/5$ has the occupancy pattern $\dots1010010100\dots$, which we denote by $\overline{23}$. 
The GSs at $\nu=2/7$ and $\nu=3/7$ are given by $\overline{34}$ and $\overline{223}$, respectively. 
Here, the numbers indicate the distance between neighboring particles, and the bar indicates the repetition of the pattern. 
In general, particles are spread out as homogeneously as possible for the given filling fraction. 

\begin{figure}
\begin{center}\includegraphics[width=0.48\textwidth]{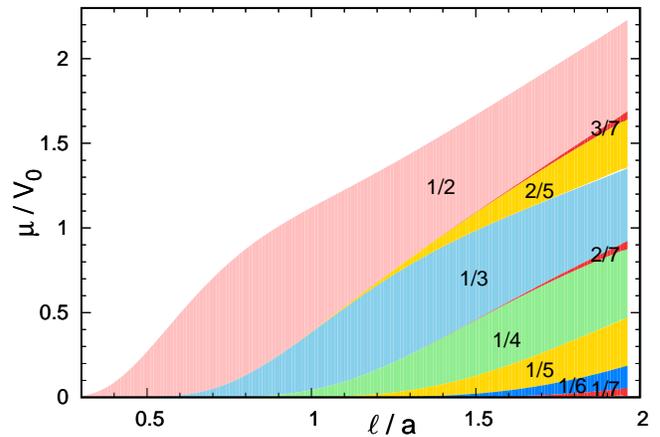}\end{center}
\caption{(color online) GS phase diagram of the effective Hamiltonian \eqref{eq:H_Vnn}. 
This effective Hamiltonian consists only of density-density interactions 
and gives an appropriate description of the original system in the limit of a large number of internal states. 
We consider the regime $\ell/a<\sigma_0\simeq 1.957$, 
where the condition of convexity is satisfied and therefore a complete devil's staircase appears. 
The number associated with each region indicates the filling factor $\nu=N/N_\phi$, 
and we focus on the case of $0\le \nu\le 1/2$. }
\label{fig:phase}
\end{figure}

The GS phase diagram of the effective Hamiltonian \eqref{eq:H_Vnn} for $\ell/a<\sigma_0$ is presented in Fig.~\ref{fig:phase}. 
We focus on the case of $0\le \nu\le 1/2$; for fermions or hard-core bosons, 
the case of $1/2<\nu\le 1$ can be treated by performing the particle-hole transformation. 
The range of each crystal GS is determined by calculating the excitation energy to a $q$-soliton state \cite{Burnell09}. 
Through such calculations, it is shown that the crystal state with $\nu=p/q$ appears for the following interval of the chemical potential: 
\begin{equation}
 \Delta\mu = \sum_{n=1}^\infty nq \left( V_{nq-1}+V_{nq+1}-2V_{nq} \right). 
\end{equation}
Notably, $\Delta\mu$ depends only on the denominator $q$ of the filling fraction, 
and decreases monotonically as a function of $q$. 
Thus, a state with smaller $q$ appears in a wider region of the parameter space as seen in Fig.~\ref{fig:phase}. 

We comment that the same type of crystal GSs are known to appear in the thin-torus limit of FQH states \cite{Tao83,Seidel05,Bergholtz07,Rotondo16}. 
In the continuous limit, the thin-torus limit and the long-range interaction in the $y$ direction in fact 
give independent routes to the same effective Hamiltonian \eqref{eq:H_Vnn}, as discussed in Appendix. 
The crystal states share a number of common features with the FQH states as follows. 
The $3$-fold degenerate crystal states $\overline{3}$ at $\nu=1/3$ in the thin-torus limit 
are known to be adiabatically connected to the topologically degenerate FQH states in the 2D case \cite{Seidel05,Nakamura10}. 
The crystal states support elementary excitations carrying a fractional charge \cite{Bergholtz07}; 
however, since such excitations appear as domain walls and have a 1D character, they do not exhibit fractional statistics. 
The crystal states also support fractional charge pumping \cite{Taddia17,Zeng16_pumping,Nakagawa17}, which is closely related to fractionally quantized conductance. 
To see it, we consider inserting a flux $\Phi$ (in units of $\hbar/Q$ with $Q$ being the fictitious charge of a particle) through the hole enclosed by a loop in the $-y$ direction. 
An adiabatic increase of the flux from $\Phi=0$ to $2\pi$ constitutes a cycle of the Hamiltonian. 
Since the inserted flux $\Phi$ modifies the boundary condition to $\Psih(x,N_y+1)=e^{i\Phi} \Psih(x,1)$, 
it leads to a translation of the Wannier orbitals by $(\Phi/2\pi) a$ in the $x$ direction. 
After the insertion of $\Phi=2\pi q$, the crystal state of period $q$ at $\nu=p/q$ moves by $q$ lattice spacings in the $x$ direction 
and recovers the original occupancy pattern, indicating pumping of $p/q$ charge quantum per cycle. 
Different fractional pumping in which a charge moves in the synthetic dimension has been discussed in Ref.~\cite{Zeng15_pumping}. 


\begin{figure}
\begin{center}\includegraphics[width=0.48\textwidth]{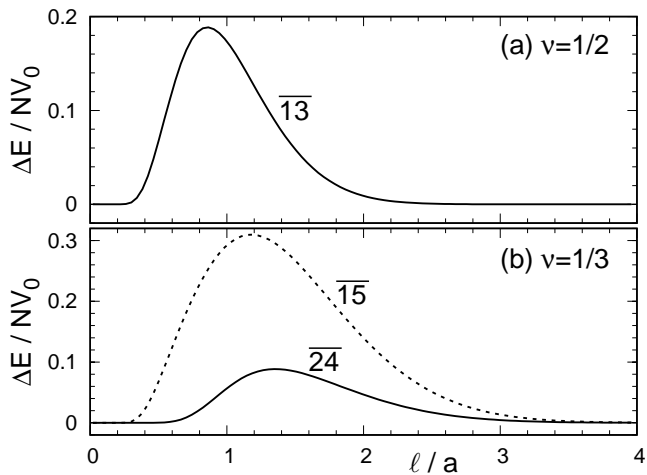}\end{center}
\caption{(color online) 
Competition of crystal states at (a) $\nu=1/2$ and (b) $\nu=1/3$. 
For $\nu=1/q$, we plot the energy of the crystal state $\overline{(q-d)(q+d)}$ (with $d=1,\dots,q-1$) 
per particle relative to that of the state $\overline{q}$, as given by Eq.~\eqref{eq:Eqd}. }
\label{fig:denergy}
\end{figure}

For $\ell/a>\sigma_0$, the condition of convexity is not fulfilled for $r\lesssim\ell/a$, 
and the most homogeneous crystal states given by Refs.~\cite{Hubbard78, Burnell09} may no longer be stable. 
For $\nu=1/2$, for example, the crystal state $\overline{2}$ can compete with another crystal state $\overline{13}$ 
(such a competition has also been discussed in a frustrated Ising model \cite{Stephenson70, Nagase76} and in the thin-torus limit of a Pfaffian quantum Hall state \cite{Bergholtz06_Pf}).
The energy of the crystal state $\overline{(q-d)(q+d)}$ (with $d=1,\dots,q-1$) per particle relative to that of the state $\overline{q}$ is given by
\begin{equation}\label{eq:Eqd}
 \frac{\Delta E}{N} = \frac12 \sum_{k=1}^\infty \left( V_{(2k-1)q-d} + V_{(2k-1)q+d} - 2V_{(2k-1)q}\right). 
\end{equation}
We plot this energy for $q=2$ and $3$ in Fig.~\ref{fig:denergy}. 
Different crystal states become asymptotically degenerate with increasing $\ell/a$. 
Thus, for $\ell/a\gtrsim 2$, the crystal GS $\overline{q}$ is easily replaced 
by other states by the effect of small perturbations such as (correlated) hopping processes present for finite $N_y$. 

\subsection{Stability against particle hopping}

We have so far considered the limit $N_y\to\infty$, and neglected the effect of particle hopping in the effective model. 
For finite $N_y$, however, hopping by a distance $N_y$ is present as in Eq.~\eqref{eq:Hkin_Jaa}, and can destabilize crystal GSs. 
Based on the energy scales of different terms in the effective Hamiltonian, 
we can predict which crystal GSs are likely to survive for finite $N_y$. 

For $N_y=2$ and $3$, we find from Eqs.\ \eqref{eq:J_Ny} and \eqref{eq:Vr}
\begin{equation}\label{eq:Vr_Ny23}
 V_0 \gg V_1 \gg J \gg V_2. 
\end{equation}
Thus, the crystal GSs at $\nu=1/2$ and $1$, which are stabilized by on-site ($V_0$) and nearest-neighbor ($V_1$) interactions, survive. 
However, other crystal GSs are likely to be destroyed by the hopping process. 

More generally, for given $N_y$, we have
\begin{equation}
 V_0 \gg V_1 \gg \cdots \gg V_{r_0} \gg J \gg V_{r_0+1},~r_0\equiv \left[ \frac{2}{\pi} N_y \right],
\end{equation}
where $[\cdot]$ is the Gauss symbol. 
We find, for example, that the crystal GS at $\nu=1/3$, which is stabilized by a combination of $V_0$, $V_1$, and $V_2$, 
survives against the hopping process for $N_y\ge 4$. 
It is essential to increase $N_y$ to stabilize crystal GSs with larger denominators of the filling fraction. 

Finally, we discuss connections of our results with Refs.~\cite{Barbarino15,Zeng15_pumping}. 
In these works, an additional optical lattice potential was introduced in 1D real space. 
It is expected, however, that our picture can qualitatively be applied 
if the magnetic flux $\phi/(2\pi)$ (in units of flux quantum) piercing through each plaquette of the synthetic square lattice is sufficiently small. 
For $\nu=1/3$ and $\phi/(2\pi)=1/4$, Zeng {\it et al.}\ \cite{Zeng15_pumping} found the emergence of density wave orders for $N_y\ge 4$, 
which is consistent with our result. 
For $N_y=3$ and $\phi/(2\pi)=1/3$, Barbarino {\it et al.}\ \cite{Barbarino15} obtained density wave orders at $\nu=1/2$, $1/3$, and $2/3$; 
here, either an infinite on-site interaction or a large nearest-neighbor interaction was introduced to stabilize these orders. 
The emergence of density wave orders at $\nu=1/3$ and $2/3$ against our analysis in Eq.~\eqref{eq:Vr_Ny23} 
can be interpreted as a consequence of an effective enhancement of $V_2$ 
due to the increased strength or range of the interaction. 

\section{Conclusions} \label{sec:summary}

In this paper, we have studied interacting bosonic or fermionic atoms in a high synthetic magnetic field in two dimensions 
spanned by continuous real space and a synthetic dimension. 
In sharp contrast with FQH systems, the interaction in such systems is highly anisotropic: 
while it is short-range in real space, it is long-range in the synthetic dimension composed of hyperfine spin states. 
Introducing an analog of the LLL approximation valid for large transition amplitudes between internal states, 
we have derived an effective 1D lattice model, 
in which particle hopping is significantly suppressed and density-density interactions play a dominant role. 
In the limit of a large number of internal states, we have shown the emergence of 
a complete devil's staircase---a crystal GS appears for every rational filling. 
In the original system, this can be understood as a consequence of the localization (delocalization) of particles 
in the $x$ ($y$) direction due to the highly anisotropic repulsive interaction in the flat LLL. 
We have also discussed the stability of crystal GSs for a small number of internal states. 


The emergence of devil's staircases has been discussed 
in dipolar gases \cite{Burnell09, Dalmonte10, Bauer12} and trapped ions \cite{Hauke10}, 
and experimentally observed in some solid-state systems \cite{Mignod77,Ohwada01,Matsuda15}. 
The emergence of crystal GSs has also been discussed in the thin-torus limit of FQH systems,
and considered as a clue for understanding basic features of FQH states 
such as topological degeneracy and fractional excitations \cite{Tao83,Seidel05,Bergholtz07,Rotondo16}. 
The present work, together with related studies \cite{Barbarino15,Zeng15_pumping}, 
demonstrates that synthetic gauge fields and highly anisotropic interactions in synthetic dimensions 
give access to this intriguing phenomenon. 

\bigskip 

The authors thank Masahito Ueda for useful discussions. 
S.\ F.\ acknowledges stimulating discussions with Eran Sela and Ce Wang 
during the 24th Annual International Laser Physics Workshop (Shanghai, August, 2015) 
and the Beijing-Tokyo Joint Workshop on Ultracold Atoms (Beijing, April, 2016). 
This work was supported by KAKENHI Grant No.~JP25800225 from the Japan Society for the Promotion of Science and by the Matsuo Foundation.

\appendix

\section{Continuum limit} \label{sec:continuum}

Throughout the paper, we have considered the situation when the $y$ coordinate is discrete as in Fig.~\ref{fig:system}. 
For comparison, we here consider the situation when both the $x$ and $y$ coordinates are continuous as in a quantum Hall problem. 
Through this comparison, it becomes clear that the expansion \eqref{eq:Psi_w} is closely analogous to the LLL approximation 
of a FQH problem on a torus geometry \cite{Yoshioka84, Haldane85}. 
Furthermore, the effective Hamiltonian consisting only of density-density interactions as in Eq.~\eqref{eq:H_Vnn} 
can be obtained by taking the thin-torus limit \cite{Tao83,Seidel05,Bergholtz07,Rotondo16} or 
by considering an artificial interaction which is long-range in the $y$ direction. 
The interaction in the latter case can be viewed as a continuum version of the spin-independent interaction \eqref{eq:Hint}. 

We consider particles of charge $Q$ on a 2D torus of size $L_x\times L_y$ subject to a uniform magnetic field $B$ in the $z$ direction (we assume $QB>0$). 
Introducing the dynamical momentum $\piv=\pv-Q\Av=-i\hbar\nabla - Q\Av$ and taking the Landau gauge $\Av=(0,Bx,0)$, 
the single-particle Hamiltonian is given by 
\begin{equation}\label{eq:qh1}
 \Kcal = \frac{\pi_x^2+\pi_y^2}{2M} 
 = \frac{\hbar\omega_c}{2} \left[ (-i\ell \partial_x)^2 + (-i\ell \partial_y-x/\ell)^2 \right],
\end{equation}
where $\ell=\sqrt{\hbar/(QB)}$ is the magnetic length and $\omega_c=QB/M$ is the cyclotron frequency. 
We require the number of flux quanta piercing the system, $N_\phi = L_xL_y/(2\pi\ell^2)$, to be an integer. 

We consider the translational symmetry of the Hamiltonian \eqref{eq:qh1}. 
Introducing the pseudomomentum 
$\Kv=\pv-Q\Av+Q\Bv\times\rv$, 
we define the translation operators in the $x$ and $y$ directions as 
\begin{equation}\label{eq:trans_xy}
\begin{split}
 &t_x (s_x) = e^{iK_x s_x /\hbar} = e^{is_x (-i\partial_x-y/\ell^2)},\\
 &t_y (s_y) = e^{iK_y s_y /\hbar} = e^{is_y (-i\partial_y)}.
\end{split}
\end{equation}
Since $\Kv$ commutes with $\piv$, it also commutes with the Hamiltonian. 
However, $t_x(s_x)$ and $t_y(s_y)$ do not in general commute with each other: 
\begin{equation}
\begin{split}
 t_x(s_x) t_y(s_y) 
 &= e^{-s_xs_y [K_x,K_y]/\hbar^2} t_y(s_y) t_x(s_x) \\
 &= e^{-is_x s_y/\ell^2} t_y(s_y) t_x(s_x). 
\end{split}
\end{equation} 
By setting, e.g., $s_x=L_x$ and $s_y=L_y/N_\phi$, 
the two translation operators become commutative. 
We can thus choose the single-particle basis where $t_x(L_x)$ and $t_y(L_y/N_\phi)$ are simultaneously diagonalized. 
The periodic boundary conditions in the $x$ and $y$ directions indicate that 
the eigenvalue of $t_x(L_x)$ must be unity while the eigenvalues of $t_y(L_y/N_\phi)$ are restricted to the $N_\phi$-th roots of unity. 

In this basis, the LLL wave functions are given by \cite{Yoshioka84}
\begin{equation}\label{eq:psi_m}
 \psi_m(\rv)=\sum_{n\in\Zbb} \frac1{\sqrt{L_y}} e^{ik_{m+nN_\phi}y} \phi (x-k_{m+nN_\phi}\ell^2) ,
\end{equation}
where $m=0,1,\dots,N_\phi-1$, $k_m=2\pi m/L_y$, and
\begin{equation}
 \phi(x)=\frac{1}{\pi^{1/4}\ell^{1/2}} e^{-x^2/(2\ell^2)}. 
\end{equation}
The wave function $\psi_m(x,y)$ is localized around $x=k_m\ell^2=mL_x/N_\phi$ with a width $\ell$ in the $x$ direction, 
and delocalized in the $y$ direction. 
Thus, these wave functions are analogous to the Wanneir orbitals shown in Fig.~\ref{fig:system}(b). 
They also satisfy the relations
\begin{equation}
\begin{split}
 &t_x(L_x/N_\phi) \psi_m (\rv) = \psi_{m+1}(\rv),\\
 &t_y(L_y/N_\phi) \psi_m(\rv) =e^{2\pi i m/N_\phi} \psi_m(\rv). 
\end{split}
\end{equation}

The many-body Hamiltonian of particles interacting via a potential $V(\rv-\rv')$ is given in the second quantized form by 
$\Hh=\Hh_\kin+\Hh_\interact$, where 
\begin{align}
 &\Hh_\kin = \int d^2 \rv  \Psih^\dagger (\rv) \Kcal \Psih (\rv), \label{eq:HQH_kin}\\
 &\Hh_\interact =\frac12 \int d^2 \rv d^2 \rv' \Psih^\dagger (\rv) \Psih^\dagger (\rv') V(\rv-\rv') \Psih(\rv')\Psih(\rv) \label{eq:HQH_int},
\end{align}
and $\Psih(\rv)$ is a bosonic or fermionic field operator. 
In the LLL approximation, the field operator $\Psih (\rv)$ is expanded as
\begin{equation}\label{eq:Psi_psi_m}
 \Psih (\rv) = \sum_{m=0}^{N_\phi-1} \ah_m \psi_m (\rv), 
\end{equation}
where $\ah_m$ is the annihilation operator for the $m$-th orbital and satisfies the commutation relations \eqref{eq:ah_comm}. 
The expansion \eqref{eq:Psi_w} for the discrete $y$ coordinate is quite analogous to Eq.~\eqref{eq:Psi_psi_m}. 
Substituting Eq.~\eqref{eq:Psi_psi_m} into Eqs.~\eqref{eq:HQH_kin} and \eqref{eq:HQH_int}, we obtain
\begin{align}
 & \Hh_\kin = \frac12 \hbar\omega_c  \sum_{m=0}^{N_\phi-1} a_m^\dagger a_m = \frac12 \hbar\omega_c \Nh, \\
 & \Hh_\interact = \frac12 \sum_{m_1,m_2,m_3,m_4} V_{m_1 m_2 m_3 m_4} a_{m_1}^\dagger a_{m_2}^\dagger a_{m_3} a_{m_4}, \label{eq:H_int_Vmmmm}
\end{align}
where 
\begin{equation}\label{eq:Vmmmm_QH}
\begin{split}
 &V_{m_1 m_2 m_3 m_4} \\
 &= \int d^2 \rv d^2 \rv' \psi_{m_1}^*(\rv) \psi_{m_2}^*(\rv') V(\rv-\rv') \psi_{m_3}(\rv') \psi_{m_4}(\rv) .
\end{split}
\end{equation}
Since $\Hh_\kin$ can be absorbed into the chemical potential, we focus on $\Hh_\interact$ hereafter. 

On a torus, the interaction potential $V(\rv-\rv')$ in general has the periodicities of the torus. 
We expand it in a Fourier series as 
\begin{equation} \label{eq:V_Vt}
 V(\rv-\rv') = \frac{1}{L_xL_y} \sum_{\qv} \Vt(\qv) e^{i\qv\cdot (\rv-\rv')},
\end{equation}
where $\qv=(\frac{2\pi}{L_x} n_x, \frac{2\pi}{L_y}n_y)~(n_x,n_y\in \Zbb)$ and
\begin{equation}\label{eq:Vq}
 \Vt (\qv) 
 = \int_0^{L_x} \int_0^{L_y} d^2 \rv e^{-i\qv\cdot\rv} V(\rv) .
\end{equation}

We calculate $V_{m_1m_2m_3m_4}$ by substituting Eqs.\ \eqref{eq:psi_m} and \eqref{eq:V_Vt} into Eq.\ \eqref{eq:Vmmmm_QH}. 
This calculation involves the integration over $\rv$ and $\rv'$ in Eq.~\eqref{eq:Vmmmm_QH}, 
the sum over $n_x$ and $n_y$ in Eq.~\eqref{eq:V_Vt}, and the sum over $n_j~(j=1,2,3,4)$ associated with $\psi_{m_j}$ in Eq.~\eqref{eq:psi_m}. 
The integration over $y$ and $y'$ lead respectively to the conditions
\begin{subequations}\label{eq:nm_condition}
\begin{align}
   n_y-m_1+m_4- n_{14} N_s &= 0, \label{eq:nm14}\\
 -n_y-m_2+m_3- n_{23} N_s &= 0 \label{eq:nm23}
\end{align}
\end{subequations}
with $n_{ij}=n_i-n_j$.  
The integration over $x$ and the sum over $n_1$ and $n_4$ are calculated as
\begin{subequations}
\begin{equation}\label{eq:int_phiphi_x}
\begin{split}
 &\sum_{n_1,n_4} \int_0^{L_x} \!\!\! dx ~\phi (x-k_{m_1+n_1N_\phi} \ell^2) \phi (x-k_{m_4+n_4N_\phi} \ell^2) e^{i q_x x} \\
 &= \sum_{n_{14}} \int_{-\infty}^\infty dx ~\phi (x-k_{m_1+n_{14}N_\phi} \ell^2) \phi (x-k_{m_4}\ell^2) e^{i q_x x} \\
 &= \sum_{n_{14}} \exp \bigg[ -\frac14 \qv^2 \ell^2  + \frac{i}2 q_x k_{m_1+m_4+n_{14}N_\phi} \ell^2 \bigg], 
\end{split}
\end{equation}
where Eq.~\eqref{eq:nm14} is used. 
Similarly, we have 
\begin{equation}
\begin{split}
 &\sum_{n_2,n_3} \int_0^{L_x} \!\!\! dx' \phi (x'-k_{m_2+n_2N_\phi}\ell^2) \phi (x'-k_{m_3+n_3N_\phi}\ell^2) e^{-i q_x x'} \\
 &= \sum_{n_{23}} \exp \bigg[ -\frac14 \qv^2 \ell^2  - \frac{i}2 q_x k_{m_2+m_3+n_{23}N_\phi}\ell^2\bigg], 
\end{split}
\end{equation}
\end{subequations}
where Eq.~\eqref{eq:nm23} is used. 
Combining these equations and rewriting the conditions \eqref{eq:nm_condition} as
\begin{subequations}\label{eq:nm_condition2}
\begin{align}
 &n_y=m_{14}+n_{14}N_s,\\
 &m_1+m_2 = m_3+m_4 - (n_{14}+n_{23})N_\phi, \label{eq:nm1234}
\end{align}
\end{subequations}
we obtain \cite{Yoshioka84}
\begin{equation}\label{eq:V_torus}
\begin{split}
 &V_{m_1 m_2 m_3 m_4}\\
 &= \frac{ \delta_{m_1+m_2,m_3+m_4}^{(N_\phi)} }{L_xL_y}  \sum_{n_x,n_y} \delta_{m_{14},n_y}^{(N_\phi)} \Vt(\qv)   \\
 &~~~~~~~\times \exp \bigg[ -\frac12 \qv^2\ell^2 + i q_x k_{m_{13}} \ell^2 \bigg] ,
\end{split}
\end{equation}
where $m_{ij}=m_i-m_j$ and $\delta_{mm'}^{(N_\phi)}$ is the Kronecker delta of period $N_\phi$ as defined in Eq.~\eqref{eq:Kronecker_p}. 

Equation \eqref{eq:V_torus} is nonzero only when $m_1+m_2\equiv m_3+m_4$ (mod $N_\phi$), 
which reflects the conservation of momentum in the $y$ direction. 
Because of this condition, the interaction Hamiltonian \eqref{eq:H_int_Vmmmm} can be rewritten as
\begin{equation}\label{eq:H_int_mn}
 \Hh_\interact = \sum_{j} \sum_{|n| \le m\le N_s/2} V_{mn}  a_{j+n}^\dagger a_{j+m}^\dagger a_{j+m+n} a_j, \\
\end{equation}
where
\begin{align}
 V_{mn} = & \frac{z_{mn}}{2} ( V_{j+n,j+m,j+m+n,j} +\epsilon_X V_{j+n,j+m,j,j+m+n} \notag\\
 &~~~+\epsilon_X V_{j+m,j+n,j+m+n,j} + V_{j+m,j+n,j,j+m+n} ), \\
 z_{mn}=&2^{-\delta_{m,|n|}(1+\delta_{m,0})} 2^{-\delta_{m,N_\phi/2}(1+\delta_{|n|,N_\phi/2})} ,\label{eq:zmn}
\end{align}
and we set $\epsilon_X=+1$ and $-1$ in the bosonic ($X=B$) and fermionic ($X=F$) cases, respectively. 
The factor $2^{-\delta_{m,N_\phi/2}(1+\delta_{|n|,N_\phi/2})}$ in Eq.\ \eqref{eq:zmn} can be ignored in the limit $N_\phi\to\infty$, 
and is dropped hereafter. 


We now take the limit $L_x/\ell\to\infty$ while keeping $L_y/\ell$ fixed. 
Then, $N_\phi$ also goes to infinity, and in Eq.~\eqref{eq:V_torus}, we can set $n_y=m_{14}$ 
and replace the sum over $n_x$ by an integral over $q_x$, obtaining
\begin{equation}\label{eq:Vjmn}
\begin{split}
  &V_{j+n,j+m,j+m+n,j}\\
 &= \frac{1}{L_y} e^{-\frac12 (k_m^2 + k_n^2)\ell^2} \int_{-\infty}^\infty \frac{dq_x}{2\pi} \Vt (q_x, k_n) e^{-\frac12 (q_x+ik_m)^2\ell^2}\\
 &\approx \frac{\Vt(-ik_m,k_n)}{\sqrt{2\pi} L_y\ell} e^{-\frac12 (k_m^2 + k_n^2)\ell^2} , 
\end{split}
\end{equation}
where we assume that $\Vt(\qv)$ is a sufficiently smooth function of $q_x$. We thus have
\begin{equation}\label{eq:Vmn_Gauss}
\begin{split}
 V_{mn} 
 &\approx \frac{z_{mn}}{2\sqrt{2\pi} L_y\ell} e^{-\frac12 (k_m^2 + k_n^2)\ell^2}\\
 &\times [\Vt(-ik_m,k_n)+ \epsilon_X\Vt(ik_n,-k_m)\\
 &~~~ + \epsilon_X\Vt(-ik_n,k_m) + \Vt(ik_m,-k_n)]  .
\end{split}
\end{equation}

In the thin-torus limit $L_y/\ell\to 0$, terms with small $m^2+n^2$ dominate the interaction Hamiltonian \eqref{eq:H_int_mn} \cite{Tao83, Seidel05,Bergholtz07,Rotondo16} 
as seen in Eq.~\eqref{eq:Vmn_Gauss}. 
Focusing on the terms with $n=0$, we obtain
\begin{equation}\label{eq:H_int_Vm0}
 \Hh_\interact = \sum_j \sum_{m\ge 0} V_{m0} : \nh_j \nh_{j+m} :,
\end{equation} 
which is analogous to Eq.~\eqref{eq:H_Vnn}. 
In particular, for a contact interaction $V(\rv-\rv')=g^\mathrm{(2D)}\delta(\rv-\rv')$ in 2D space for bosons, the coefficient $V_{m0}$ is given by
\begin{equation}
 V_{m0} = \frac{2g^\mathrm{(2D)}}{4^{\delta_{m0}}\sqrt{2\pi} L_y\ell} e^{-\frac12 k_m^2 \ell^2}.
\end{equation}

Alternatively, a similar Hamiltonian can also be obtained by 
considering an artificial interaction $V(\rv-\rv')=g\delta(x-x')$, 
which is short- and long-range in the $x$ and $y$ directions, respectively. 
This interaction can be viewed as a continuum version of the spin-independent interaction \eqref{eq:Hint}. 
By substituting its Fourier transform $\Vt(\qv)=g L_y \delta_{n_y,0}$ into Eq.~\eqref{eq:V_torus} 
and take the limit $L_x/\ell\to\infty$ as in Eq.~\eqref{eq:Vjmn}, we obtain
\begin{equation}\label{eq:V_torus_g}
 V_{m_1m_2m_3m_4} = \delta_{m_1m_4}\delta_{m_2m_3} \frac{g}{\sqrt{2\pi}\ell} e^{-\frac12 k_{m_{13}}^2 \ell^2}. 
\end{equation}
The interaction Hamiltonian then has the same form as Eq.~\eqref{eq:H_int_Vm0}, 
where the coefficients $V_{m0}$ are replaced by
\begin{equation}\label{eq:Vm0}
 V_{m0} = \frac{g}{2^{\delta_{m,0}}\sqrt{2\pi} \ell} e^{-\frac12 k_m^2 \ell^2} .
\end{equation} 
In this case, notably, we do not need to take the thin-torus limit $L_y/\ell\to 0$ to arrive at the simple Hamiltonian \eqref{eq:H_int_Vm0} 
that consists only of density-density interactions. 
Thus, the thin-torus limit $L_y/\ell\to 0$ and the long-range interaction in the $y$ direction 
provide independent routes to the Hamiltonian \eqref{eq:H_int_Vm0}. 
If we introduce $a\equiv L_x/N_\phi = 2\pi \ell^2/L_y$ (the spacing between neighboring LLL orbitals in the $x$ direction), 
we have $k_m \ell = am/\ell$ and thus find the direct correspondence between Eq.\ \eqref{eq:Vr} and Eq.\ \eqref{eq:Vm0}.



\begin{thebibliography}{99}
\bibitem{Comment_dressed}
The calculation of dressed states is equivalent to solving the tight-binding model on a finite chain of $N_y$ sites. 
Here, the nearest-neighbor hopping involves the Peierls phase factor $e^{\pm ibx}$ for an atom at $x$ as seen in Fig.~\ref{fig:system}(a). 
For open boundary conditions in the synthetic dimension, 
this phase factor can be gauged away in the atomic problem, leading to spatially-independent dressed levels. 
For a periodic boundary condition, in contrast, 
a magnetic flux of $N_y bx$ pierces through the effective ``ring'' formed by the internal states, 
leading to spatially-dependent dressed levels as shown by the effective cosine potentials in Fig.~\ref{fig:system}(b).
\end{thebibliography}
\end{document}